\documentclass[aps,prd,twocolumn,superscriptaddress,nofootinbib]{revtex4-1}
\usepackage{graphicx,color}
\usepackage{latexsym,amsmath,amssymb,graphicx,booktabs}
\usepackage{hyperref}
\usepackage{float}

\usepackage{scalerel,stackengine}
\stackMath
\newcommand\reallywidehat[1]{%
\savestack{\tmpbox}{\stretchto{%
  \scaleto{%
    \scalerel*[\widthof{\ensuremath{#1}}]{\kern-.6pt\bigwedge\kern-.6pt}%
    {\rule[-\textheight/2]{1ex}{\textheight}}%WIDTH-LIMITED BIG WEDGE
  }{\textheight}% 
}{0.5ex}}%
\stackon[1pt]{#1}{\tmpbox}%
}
\def\VEV#1{{\left\langle #1 \right\rangle}}

\newcommand{\SNR}{{\mathrm{SNR}}}

\def\cleb#1#2#3#4#5#6{\langle #3 \, #4 \, #5 \, #6 | #1#2 \rangle}
\def\ALMt#1#2#3#4{A^{#1 #2}_{#3 #4}}
\def\ALM{\ALMt{L}{M}{\ell}{\ell'}}

\graphicspath{ {./Graphs/} }

\begin{document}

\title{Chirality of the gravitational-wave background and
pulsar-timing arrays}

\author{Enis Belgacem}
\email{enis.belgacem@unige.ch}
\affiliation{D\'epartement de Physique Th\'eorique and Center for Astroparticle Physics, Universit\'e de Gen\`eve, 24 quai Ansermet, CH--1211 Gen\`eve 4, Switzerland}

\author{Marc Kamionkowski}
\email{kamion@jhu.edu}
\affiliation{Department of Physics and Astronomy, Johns Hopkins University, 3400 N. Charles St., Baltimore, MD 21218, USA}

\begin{abstract}
We describe the signatures of a circularly polarized
gravitational-wave background on the timing residuals
obtained with pulsar-timing arrays.  Most generally, the
circular polarization will depend on the gravitational-wave
direction, and we describe this angular dependence in terms of
spherical harmonics.  While the amplitude of the monopole (the
overall chirality of the gravitational-wave background) cannot
be detected, measures of the anisotropy are theoretically conceivable.  We provide
expressions for the minimum-variance estimators for the
circular-polarization anisotropy.  We evaluate the smallest
detectable signal as a function of the signal-to-noise ratio with
which the isotropic GW signal is detected and the number of
pulsars (assumed to be roughly uniformly spread throughout the
sky) in the survey.  We find that the overall dipole of the circular polarization and a few higher overall multipoles, are detectable in a survey with $\gtrsim100$ pulsars if their amplitude is close to
maximal and once the isotropic signal is established with a
signal-to-noise ratio $\gtrsim400$.  Even if the anisotropy can be established, though, there will be limited information on its direction.  Similar arguments apply to astrometric searches for gravitational waves.
\end{abstract}

\maketitle

\section{Introduction}
\label{sec:intro}
A gravitational wave passing between the Earth and a pulsar is
known to affect the periodicity of the observed pulses
\cite{Detweiler:1979wn,Sazhin:1978}. The effect can be encoded
in the timing residual, defined as the relative difference
between the observed period of pulses and the one produced by
the pulsar.  The explorable frequency range 
roughly goes from a few nHz to 1 $\mu$Hz, the lower limit being
determined by the time span of observations and the upper limit
by the data sampling rate. Monitoring and correlating the
irregularities in the signals emitted by different pulsars
allows an indirect study of gravitational waves (GWs) and has
led to the idea of pulsar timing arrays (PTAs)
\cite{Foster:1990,Maggiore:1999vm,Burke-Spolaor:2015xpf,Lommen:2015gbz,Hobbs:2017oam,Yunes:2013dva,Hobbs:2013aka,Manchester:2012za,Arzoumanian:2018saf,Lentati:2015qwp,Verbiest:2016vem}
to detect gravitational waves at $\sim$nHz--$\mu$Hz frequencies. In
particular it may be possible to extract information on the stochastic
gravitational-wave background due to supermassive-black-hole (SMBH)
mergers \cite{Rajagopal:1994zj,Jaffe:2002rt}.  There are also prospects
to augment PTA measurements with information from stellar astrometry
\cite{Book:2010pf, Moore:2017ity, Mihaylov:2018uqm,
OBeirne:2018slh, Qin:2018yhy}.

A stochastic background from SMBH mergers may well be
anisotropic, given the uneven distribution of SMBH mergers on
the sky
\cite{Allen:1996gp,Sesana:2008xk,Ravi:2012bz,Cornish:2013aba,Kelley:2017vox}
and prior work
\cite{Anholm:2008wy,Mingarelli:2013dsa,Gair:2014rwa,Hotinli:2019tpc}
has developed tools to seek and characterize anisotropies in the
intensity of the GW background  with PTAs/astrometry.  However,
GWs from SMBH mergers will most generally be circularly
polarized.  Therefore, the stochastic
GW background is likely to be circularly polarized, with an
amplitude that varies across the sky.  Ref.~\cite{Kato:2015bye}
discussed techniques to seek this circular polarization with
PTAs.

In this paper we re-visit the PTA search for circular
polarization with a simple augmentation of recent work
\cite{Hotinli:2019tpc} on the detection of angular GW-intensity
fluctuations.  Unlike most prior related work,
Ref.~\cite{Hotinli:2019tpc} discussed angular fluctuations in
harmonic space, rather than configuration space, an alternative
approach that provides elegant/economical mathematical
expressions, simple estimates for signal detectability, and some
novel insights.  Here, we show how that work is easily altered
to allow a search for circular polarization.  While the results
are formally equivalent to what was presented in
Ref.~\cite{Kato:2015bye}, the formalism here allows for
more compact mathematical expressions and some associated
insights.

Ref.~\cite{Hotinli:2019tpc} idealized measurements of a
timing-residual $z(\hat n,t)$ as a function of position $\hat n$
on the sky and time $t$.  The time dependence was then described
in terms of its Fourier amplitudes for frequency $f$ (one real
amplitude for the sine, with respect to some nominal $t=0$ time,
and another for the cosine for each $f$),  The resulting
Fourier maps $z_f(\hat n$) were then decomposed in terms of
spherical-harmonic coefficients $z_{f,\ell m}$.  Estimators for
angular intensity fluctuations were then constructed from
bipolar spherical harmonics (BiPoSHs)
\cite{Hajian:2003qq,Hajian:2005jh,Joshi:2009mj}, and in
particular from BiPoSHs of even parity.

In this paper, estimators for the circular polarization of the
GW background will be similarly constructed but with a few
notable differences:  First, a circularly polarized GW is a
linear combination of two linear polarizations that are out of
phase.  Thus, a circular-polarization estimator requires that we
consider the sine and cosine amplitudes {\it together} for any given
frequency $f$, which we do here by allowing $z_f(\hat n)$ to be
complex.  We then show that circular-polarization estimators look
identical to those for intensity fluctuations, but for {\it
odd}-parity (rather than even-parity) BiPoSHs
\cite{Book:2011na}.

The plan of the paper is as follows:  We review in
Section \ref{sec:expansion} the expansion of the timing
residuals in terms of spherical harmonics and review the
BiPoSH formalism that will be used to
construct estimators for circular-polarization anisotropies.
Section \ref{sec:background} presents the model we assume for
the stochastic background and obtains predictions for the
observables for this background.  Section \ref{sec:estimators}
presents the estimators for the circular-polarization
anisotropies, and formal expressions for the variances with
which these estimators can be measured.  Section
\ref{sec:result} then presents quantitative results for the
smallest detectable circular-polarization anisotropies, and
concluding remarks are presented in Section
\ref{sec:conclusions}.

\section{Spherical-harmonic expansion and bipolar spherical
harmonics (BiPoSHs)}
\label{sec:expansion}

We imagine a set of pulsars spread roughly uniformly across the
sky so that the GW-induced timing residual $z(\hat n,t)$ can be
obtained as a function of time $t$ and position $\hat n$.  The
time sequence can then be represented equivalently in terms of
the Fourier components $z_f(\hat n)$ for frequency $f$, and the
angular pattern can then be represented in terms of the
spherical-harmonics components as
\begin{equation}
\label{eq:defzlm}
z_f(\hat{n})= \sum_{\ell=0}^\infty \sum_{m=-\ell}^L z_{f,\ell m}
Y_{\ell m}(\hat n).
\end{equation}
In Ref.~\cite{Hotinli:2019tpc}, it was presumed that $z_f(\hat
n)$ could be taken to be real: the intensity-fluctuation
analysis therein could be performed independently on either the
real or the imaginary part (or equivalently, on the amplitudes
of the cosine or sine of any particular $f$ mode).  Thus, in
that work (as in work on CMB temperature fluctuations), we had
$z_{f,\ell m}^*= (-1)^m z_{f,\ell,-m}$.  Put another way, the
$2\ell+1$ independent coefficients for any given $\ell$ could be
taken to be $z_{f,\ell0}$, the $\ell$ real parts $\sqrt{2}\, {\rm
Re}\,z_{f,\ell m}$ of $z_{f,\ell m}$ for $m>0$, and the $\ell$
imaginary components $\sqrt{2}\, {\rm Im}\, z_{f,\ell m}$ for
$m>0$.

For the
analysis here, however, the complexity of $z_f(\hat n)$---i.e.,
the relative amplitudes of the cosine and sine mode for a given
$f$---is essential.  Thus, in this paper, $z_f(\hat n)$ is most
generally complex, and so $z_{f,\ell m}^*$ is not necessarily
equal to $(-1)^m z_{f,\ell m}$.  For any given $\ell$, there are
now $2(2\ell+1)$ components of $z_{f,\ell m}$ which can be taken
to be the real and imaginary parts for all $-\ell \leq m \leq
\ell$.

Also, for notational economy, we suppress below the
subscripts $f$ on the map, the spherical-harmonic coefficients,
and power spectra.  It should be understood that throughout the
rest of the paper, it is assumed that the analysis is done for
this one frequency component $f$.  We then discuss in the
Conclusions how to incorporate multiple frequencies.

A model for the stochastic background makes no predictions for
the specific values of $z_{\ell m}$.  Rather, it makes
predictions for their correlations.  
The most general two-point correlation between any two
$z_{\ell m}$ takes the form (see, e.g.,
Refs.~\cite{Pullen:2007tu,Book:2011na}),
\begin{eqnarray}
\label{eq:biposh}
     \VEV{ z_{\ell m} z^*_{\ell'm'}} &=& C_{\ell} \delta_{\ell\ell'}
     \delta_{mm'} \nonumber \\
     & & + \sum_{L=1}^\infty\sum_{M=-L}^L  (-1)^{m'}
     \cleb{L}{M}{\ell}{m}{\ell',}{-m'} \ALM,\nonumber \\
\end{eqnarray}
where the $\ALM$ are bipolar spherical harmonics (BiPoSH)
coefficients
\cite{Hajian:2003qq,Hajian:2005jh,Joshi:2009mj}.  If
the stochastic background is statistically isotropic and
unpolarized, then $A^{LM}_{\ell\ell'}=0$ for all $L\geq1$.
Ref.~\cite{Hotinli:2019tpc} found that anistropies in the
intensity of the GW background resulted in nonzero BiPoSH
coefficients of {\it even} parity (i.e., $L+\ell+\ell'=$even)
only.  We will see that circular polarization induces {\it
odd}-parity BiPoSHs, those with $L+\ell+\ell'=$odd.

\subsection{Estimators of BiPoSH coefficients}

The measured timing-residual coefficients are assumed to be
$z_{f,\ell m}^{\rm data}=z_{f,lm}+z_{f,\ell m}^{\rm noise}$ with
\begin{equation}
     \VEV{ z_{f,\ell m}^{\rm noise} z_{f,\ell'm'}^{*\,{\rm noise}}} =
     N_f^{zz} \delta_{\ell\ell'}\delta_{mm'},
\end{equation}
with the noise power spectrum $N_f^{zz}$ independent of $\ell$
(as will arise in the idealized scenario of pulsars distributed
roughly uniformly on the sky, with comparable timing noises).
The BiPoSH coefficients are estimated from data as
\begin{equation}
     \reallywidehat{\ALM} = \sum_{mm'} z^{\rm data}_{lm} z^{*\,{\rm
     data}}_{l'm'} (-1)^{m'} \cleb{L}{M}{l}{m}{l',}{-m'}.
\label{eq:ALMestimator}
\end{equation}
The variance of this estimator was evaluated under the null
hypothesis of a Gaussian and isotropic
map, in Ref.~\cite{Book:2011na}.  That analysis assumed,
however, a real map, whereas we are now taking $z(\hat n)$ to
be complex.  As a result there is no requirement for
$A^{LM}_{\ell\ell'}$ to be antisymmetric (for $L+\ell+\ell'=$odd)
under $\ell\leftrightarrow \ell'$ nor for the
$A^{LM}_{\ell\ell}$ to vanish for $\ell+\ell'+L$=odd.  For a
complex map and for $L+\ell+\ell'$=odd, 
\begin{equation}
     \VEV{ \left|\reallywidehat{\ALM} \right|^2} =
     C_\ell^{\mathrm{data}} C_{\ell'}^{\mathrm{data}},
\label{eq:ALMvariance}
\end{equation}
where $C_\ell^{\mathrm{data}} = C_\ell + N^{zz}$ includes both
the signal and the noise power spectra.

The estimator for the isotropic power spectrum $C_\ell$ is
\begin{equation}
     \reallywidehat{C_\ell} = \sum_{m=-\ell}^\ell \frac{ |z_{\ell
     m}^{\rm data}|^2}{2\ell+1} - N^{zz},
\end{equation}
and its variance is
\begin{equation}
\label{eq:varCell}
     \VEV{ \left(\Delta C_\ell \right)^2} = \frac{1}{2\ell+1}
     \left(C_\ell^{\rm data} \right)^2.
\end{equation}
Note that this expression differs from that, more commonly seen,
for the case where $z(\hat n)$ is real.  As discussed above, in
that case, each $C_\ell$ is estimated from the $2\ell+1$
independent components of $z_{\ell m}$.  If $z(\hat n)$ is
complex, though, then there are $2 (2\ell+1)$ independent
components of $z_{\ell m}$ yielding a replacement $2\ell+1
\rightarrow 2(2\ell+1)$ relative to the more familiar equation.

\section{A polarized background and its timing residuals}
\label{sec:background}

\subsection{Spherical-harmonic coeffcients}

Eq.~(18) in Ref.~\cite{Hotinli:2019tpc} provides the
spherical-harmonic coefficients, induced by a single
gravitational wave of frequency $f$ propagating in the $\hat
k$ direction.  Identifying the GW circular-polarization
amplitudes $h_R=2^{-1/2} (h_+ + i h_\times)$ and $h_L=2^{-1/2} (h_+ - i
h_\times)$ in terms of the linear-polarization amplitudes $h_+$
and $h_\times$, that expression can be written \footnote{We correct here a missing factor of two in
Eqs. (14) and (15) of Ref.~\protect\cite{Hotinli:2019tpc} (that does not
affect the final quantitative results for anisotropy estimators
presented there in terms of the signal-to-noise ratio SNR with
which the isotropic signal is detected.)}
\begin{equation}
     z_{\ell m}(\hat k) = 2^{-1/2} z_\ell  \left[ h_L D^{(\ell)}_{m2} +
     h_R D^{(\ell)}_{m,-2} \right],
\end{equation}
where $ D^{(\ell)}_{mm'}(\phi_k,\theta_k,0)$ are the Wigner
rotation functions specified by the three Euler angles $\phi_k$,
$\theta_k$ and $\psi_k=0$ in the $z$-$y$-$z$ convention, and
\begin{equation}
\label{eq:zell}
     z_\ell \equiv (-1)^{\ell}
     \sqrt{\frac{{4\pi}(2\ell+1) (\ell-2)!}{(\ell+2)!}}.
\end{equation}
Only harmonics coefficients with $\ell \geq 2$ are generated in the
timing residuals map.

The most general gravitational-wave background is then described
by a superposition of plane waves propagating along any
direction $\hat k$.  The spherical-harmonic coefficients for
this background are then
\begin{equation}
\label{eq:finalzlm}
     z_{\ell m}(\hat k) = 2^{-1/2} z_\ell \int \frac{d^3k}{(2\pi)^3} \left[
     h_L(\vec k) D^{(\ell)}_{m2}(\vec k) +h_R(\vec k) D^{(\ell)}_{m,-2}(\vec k)
     \right], 
\end{equation}
where we have summed over all GW wavevectors $\vec k = 2 \pi f
\hat k$.

\subsection{A circularly polarized gravitational-wave background}

\label{subsec:model}
We now consider a gravitational-wave background described by the
following wave-amplitude correlations:
\begin{align}
\label{eq:background}
     \VEV{ h_R(\vec k) h_{R}^*(\vec k') } &= \frac 14 
     (2\pi)^3 \delta_D(\vec k -\vec k') P_h(k) \left[1-\epsilon (\hat k)\right] , \nonumber \\
       \VEV{ h_L(\vec k) h_{L}^*(\vec k') } &= \frac 14 
     (2\pi)^3 \delta_D(\vec k -\vec k') P_h(k) \left[1+\epsilon (\hat k)\right],\nonumber \\
 \VEV{ h_R(\vec k) h_{L}^*(\vec k') } &= 0 .
\end{align}
The ``chirality function'' $\epsilon (\hat{k})$ is assumed to
depend only on the direction of propagation, and it parametrizes the
degree of circular polarization for GWs moving in direction
$\hat k$.  It can be decomposed in spherical harmonics as
\begin{equation}
     \epsilon (\hat{k})= \sum_{L=0}^\infty \sum_{M=-L}^L
     \epsilon_{LM} Y_{LM}(\hat k).
\end{equation}
Comparing to Eq. (13)
of~\cite{Hotinli:2019tpc}, where the index $L$ is constrained to
assume strictly positive values (i.e., $L\geq1$), here the $L=0$
term is in principle allowed because it cannot be reabsorbed into a
redefinition of $P_h(k)$ for both the right-handed and
left-handed power spectra in Eq.~(\ref{eq:background}). However,
we will see that the monopole gives no contribution to the
correlators of timing residuals and is therefore not
detectable.

Positivity of power spectra imposes the restrictions
$\epsilon_{L0} \leq \sqrt{4\pi/(2L+1)}$ (with similar bounds for
$\sqrt{2}\,\mathrm{Re}\, \epsilon_{LM}$ and
$\sqrt{2}\,\mathrm{Im}\, \epsilon_{LM}$ for $M\neq0$).
The correlators between timing residual-coeffcients are then
evaluated using Eq.~(\ref{eq:finalzlm}) and are given by
\begin{multline*}
 \label{eq:integral} 
     \VEV{z_{\ell m} z_{\ell'm'}^*} = \frac 18 z_\ell z_{\ell'} \int 
  \frac{d^3k}{(2\pi)^3} P_h(k)  \Bigg\{ D^{(\ell)}_{m2}(\hat k) \left(D^{(\ell')}_{m'2}(\hat k) \right)^* \\
\times \left[1+\sum_{LM}\epsilon_{LM} Y_{LM}(\hat k)\right]+D^{(\ell)}_{m,-2}(\hat k) \left(D^{(\ell')}_{m',-2}(\hat k) \right)^* \\
\times \left[1-\sum_{LM}\epsilon_{LM} Y_{LM}(\hat k)\right]\Bigg\}.
\end{multline*}
The integration over directions $\hat{k}$ leads to correlators
of the form in Eq.\ (\ref{eq:biposh}) with 
\begin{equation}
     C_\ell = \frac{z_\ell^2}{4(2\ell+1)} I,
\label{eq:powerspectrum}     
\end{equation}
and
\begin{equation}
     A^{LM}_{\ell\ell'} =\frac{1-(-1)^{\ell+\ell'+L}}{2}
     (-1)^{\ell-\ell'}(4\pi)^{-1/2} \epsilon_{LM} \frac14 z_\ell
     z_{\ell'} H^L_{\ell\ell'} I,
\label{eq:biposhresult}     
\end{equation}
where
\begin{equation}
     H^L_{\ell\ell'} \equiv \left( \begin{array}{ccc} \ell & \ell' &
     L \\ 2 & -2&  0 \end{array} \right),
\end{equation}
and ${I \equiv  [4\pi/(2\pi)^3] \int k^2\,dk\,
P_h(k)}$.

The coefficient $\left[1-(-1)^{\ell+\ell'+L}\right]/2$ selects
only odd-parity BiPoSHs (odd values of $\ell+\ell'+L$).
Comparing to Eq.~(22) of Ref.~\cite{Hotinli:2019tpc} and the
discussion therein, we see that the correlations induced by
circular-polarization anisotropies differ from those of
intensity anisotropies in the parity of the BiPoSHs allowed (odd
for circular polarization and even for intensity).  The other
difference is that $A^{LM}_{\ell\ell'}$ is not degenerate here
with $A^{LM}_{\ell'\ell}$ (as it is for the intensity
estimator), as $z(\hat n)$ here is taken to be the complex sum
of the amplitudes of the sine and cosine of the frequency $f$.

\section{Chirality estimators}
\label{sec:estimators}

Estimators for the chirality coefficients $\epsilon_{LM}$ are
obtained in direct analogy with Ref.~\cite{Hotinli:2019tpc} for
a survey parametrized by the signal-to-noise ratio (SNR) with
which the isotropic GW background is detected and the maximum
multipole moment $\ell_{\rm max}$ (which is $\ell_{\rm max}\sim
\sqrt {N_p}$ for a sky map with $N_p$ pulsars  distributed
roughly uniformly on the sky) accessible with the survey.
The SNR is obtained by summing in quadrature the
SNRs for each accessible multipole $\ell$, assuming an error on
$C_\ell$ given by Eq.~(\ref{eq:varCell}) with $C_\ell^{\rm
data}=N^{zz}$, corresponding to the null hypothesis of no
gravitational-wave background.  The resulting SNR for the
frequency channel $f$ is
\cite{Roebber:2016jzl,Hotinli:2019tpc} (noting the extra factor
of 2 for the complexity of $z(\hat n)$),
\begin{equation}
  \SNR =\left[ \sum_{\ell=2} ^{\ell_{\rm max}} (2\ell+1) \left(
  \frac{C_\ell}{N^{zz}}\right)^2 \right]^{1/2} \simeq \frac{\pi I}{6\sqrt
  3 N^{zz}}.
\label{eq:I}
\end{equation}
The approximation holds for any $\ell_{\rm max}$ given that the
sum is dominated very heavily by the lowest-$\ell$ terms.

Following the analysis in Ref.~\cite{Hotinli:2019tpc}, the
minimum-variance estimator for each chirality amplitude
$\epsilon_{LM}$ is
\begin{equation}
\label{eq:epsLM}
     \reallywidehat{\epsilon_{LM}} = \frac{\sum_{\ell\ell'} (\reallywidehat{\epsilon_{LM}})_{\ell
     \ell'}  (\Delta \epsilon_{LM})_{\ell\ell'}^{-2}}{
     \sum_{\ell\ell'} (\Delta \epsilon_{LM})_{\ell\ell'}^{-2}},
\end{equation}
where
\begin{equation}
\label{eq:epsestimator}
    (\reallywidehat{\epsilon_{LM}})_{\ell \ell'} =
    (-1)^{\ell-\ell'} 4 \sqrt{4\pi}\frac{
    \reallywidehat{A^{LM}_{\ell\ell'}}}{z_\ell z_{\ell'}
    H^L_{\ell \ell'} I},
\end{equation}
is the contribution of each $\ell\ell'$ pair to the estimator,
and
\begin{eqnarray}
     (\Delta \epsilon_{LM})_{\ell\ell'}^2
     &=& \frac{64 \pi  C_\ell^{\rm data}
     C_{\ell'}^{\rm data}}{(z_\ell z_{\ell'}
     H^L_{\ell\ell'} I)^2} \nonumber \\
     &=& \frac{16\pi^3}{27} \frac{ C_\ell^{\rm
     data} C_{\ell'}^{\rm data}}{ \left(z_\ell z_{\ell'}
     H^L_{\ell\ell'} \right)^2 (\SNR)^2(N^{zz})^2},
\end{eqnarray}
is the variance of each of these contributions.  The variance of
the combined estimator $\reallywidehat{\epsilon_{LM}}$ is then
\begin{equation}
\label{eq:varepsLM}
    (\Delta \epsilon_{LM})^{-2}=\sum_{\ell\ell'} (\Delta
    \epsilon_{LM})_{\ell\ell'}^{-2}.
\end{equation}
It is independent of $M$.  The $\ell,\ell'$ sums here are over
$\ell+\ell'+L$=odd.  Since the $A^{LM}_{\ell\ell'}$ are not
necessarily antisymmetric in $\ell,\ell'$ (since $z(\hat n)$ is
not real here), the sums are over all $\ell,\ell'$ pairs (not
just those with $\ell'\geq \ell$ as in
Ref.~\cite{Hotinli:2019tpc}).

Using Eqs.~(\ref{eq:powerspectrum}) and (\ref{eq:I}) we can express
Eq.~(\ref{eq:varepsLM}) in terms of the SNR for the detection of
the isotropic unpolarized signal and, furthermore, the noise
power spectrum $N^{zz}$ cancels out from the final result,
leaving us with 
\begin{eqnarray}
\label{eq:finalvarepsLM}
      (\Delta
      \epsilon_{LM})^{-2}&=&\frac{27}{16\pi^3}\sum_{\ell\ell'}
      \left(z_\ell z_{\ell'}H^L_{\ell\ell'} \right)^2
      \left(\frac{1}{\rm{SNR}}+\frac{3\sqrt 3}{2\pi} 
      \frac{z_\ell^2}{2\ell+1}\right)^{-1} \nonumber \\
      & & \times
      \left(\frac{1}{\rm{SNR}}+\frac{3\sqrt 3}{2\pi}
      \frac{z_\ell'^2}{2\ell'+1}\right)^{-1}.
\end{eqnarray}
This expression can then be evaluated for any nominal SNR with
which the isotropic signal is detected and taking the sums up to
$\ell_{\rm max} \sim N_p^{1/2}$, with $N_p$ the number of
pulsars.

This expression evaluates, in the limit $\SNR \to \infty$, to
\begin{equation}
      (\Delta \epsilon_{LM})^{-2} \to \frac{1}{4\pi}
      \sum_{\ell\ell'} (2\ell+1)(2\ell'+1) \left(
      H^L_{\ell\ell'} \right)^2,
\end{equation}
and in the limit $\SNR \to 0$ to
\begin{equation}
      (\Delta \epsilon_{LM})^{-2} \to \frac{27}{16\pi^3} \SNR^2
      \sum_{\ell\ell'} \left(z_{\ell} z_{\ell'} H^L_{\ell\ell'}
      \right)^2.
\end{equation}

\section{Results}
\label{sec:result}

\subsection{Monopole is not observable}

As anticipated in Section~\ref{sec:background}, the monopole
term $\epsilon_{00}$ is not observable.  This is because the
$L=M=0$ BiPoSH coefficients, $A^{00}_{\ell\ell} =
(-1)^\ell \sqrt{2\ell+1} C_\ell$, all have $\ell=\ell'$
and therefore always have $L+\ell+\ell'=$even.  This agrees with
a similar conclusion in Ref.~\cite{Kato:2015bye} obtained using
overlap reduction functions.

\subsection{Dipole anisotropy}
\label{sec:dipoleresult}
We now consider the lowest observable multipole, the dipole
$L=1$.  Given the triangle constraint $|\ell-\ell'| \leq L$ and
$\ell+\ell'+L$=odd, only $\ell=\ell'$ contributes to the
sum.  We then use $(H^1_{\ell\ell})^2 = 4\left[\ell(\ell+1)(2\ell+1)
\right]^{-1}$ to obtain for the smallest detectable (at
$3\sigma$) signal,
\begin{eqnarray}
\label{eq:deltaeps1M}
     \epsilon_{1M,{\rm min}}&=&3 \Delta \epsilon_{1M} =\frac{1}{2}
     \sqrt{\frac{\pi}{3}} \left \{
     \sum_{\ell=2}^{\ell_{\rm max}} \frac{2\ell+1}{\left[(\ell+2)(\ell-1)\right]^{2} \left[ \ell(\ell+1) \right]^3} \right. \nonumber
     \\
     & &  \times \left. \left[\frac{1}{\rm{SNR}}+6\sqrt 3 \frac{(\ell
     -2)!}{(\ell
     +2)!}\right]^{-2} \right\}^{-1/2}.   
\end{eqnarray}

We can understand this result analytically by considering the
asymptotic behaviors in the limits of high and low
signal-to-noise.  When $\SNR\to\infty$ and $\ell_{\rm max}\gg 1$
the sum in Eq.~(\ref{eq:deltaeps1M}) converges to \footnote{Here, $\gamma_E$ is Euler's constant and the finite correction $2\gamma_E-5/2\simeq-1.35$ is relevant as it gives a 13.6\% correction on $\epsilon_{1M,{\rm min}}$ for $\ell_{\rm max}=20$.}
\begin{equation}
   \epsilon_{1M,{\rm min}} \simeq \frac{3 \sqrt{\pi}}{\sqrt{2\ln
   \ell_{\rm max}+2\gamma_E-5/2}}, \quad {\rm   as}\,~\SNR\to\infty.
\label{eqn:highSNR}   
\end{equation}
In the low-SNR limit we find
\begin{equation}
   \epsilon_{1M,{\rm min}} \simeq \frac{13.2}{\SNR}, \quad {\rm
   as}\,~\SNR\to 0.
\label{eqn:lowSNR}   
\end{equation}
\begin{figure}[h]
\centering
\hspace*{-0.1cm}\includegraphics[width=0.5\textwidth]{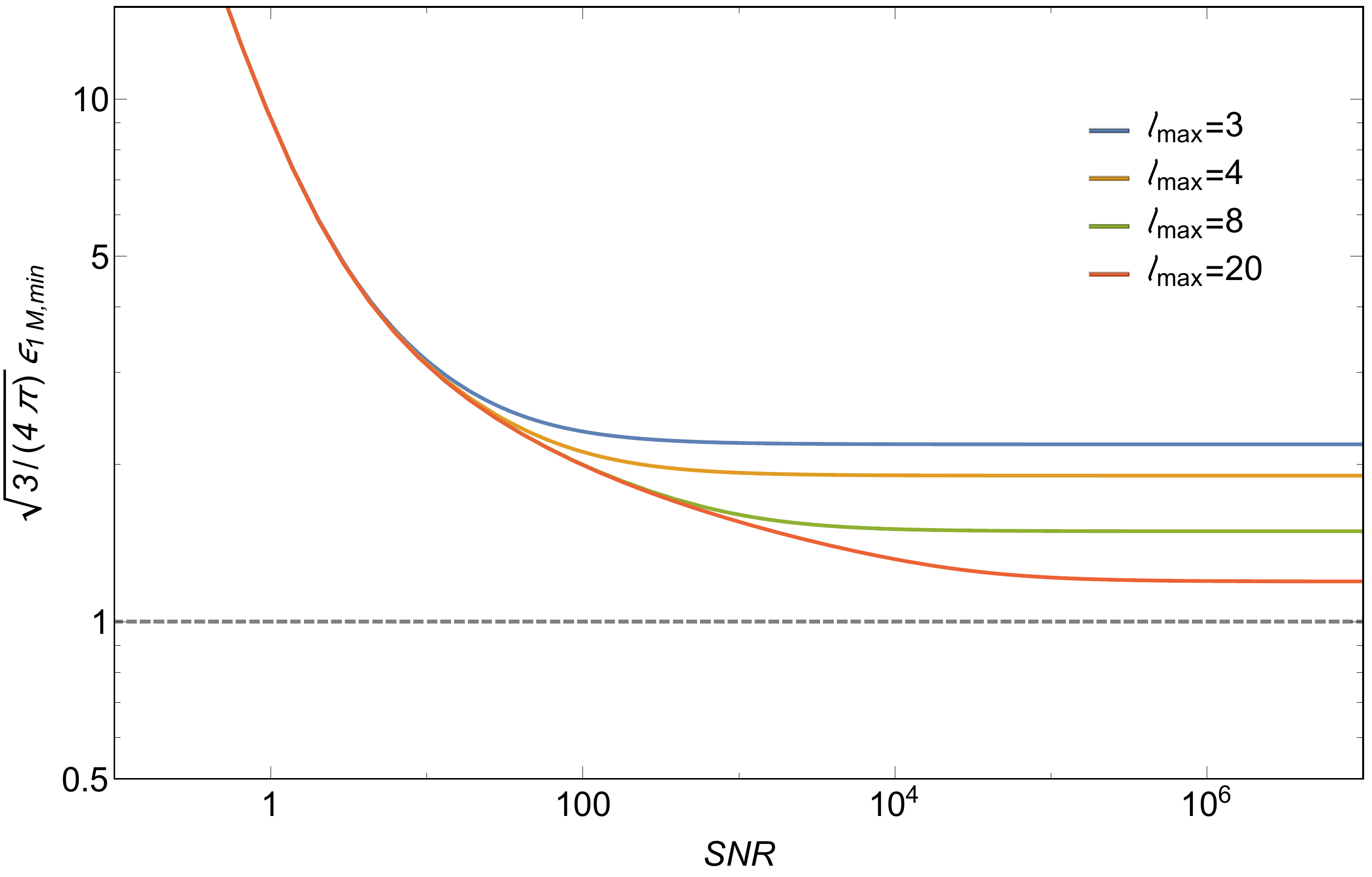}
\caption{The smallest detectable (at the $3\sigma$ level)
circular-polarization dipole $\epsilon_{1M}$ (normalized to its maximum allowed value $\epsilon_{1M,{\rm max}} =
\sqrt{4\pi/3}$), as a function of the SNR with which the isotropic GW background is detected, for several values of the maximum timing-residual multipole moment $\ell_{\rm max}$.}
\label{fig:dipole}
\end{figure}
 Fig.~\ref{fig:dipole} shows the smallest detectable dipole coefficient $\epsilon_{1M,{\rm min}}$ as a function of the isotropic signal SNR, for a few values of $\ell_{\rm max}$. The value $(\rm SNR)_{\rm high}$ at which $\epsilon_{1M,{\rm min}}$ reaches its aymptotic value can be evaluated by looking at the second line of Eq.~\ref{eq:deltaeps1M} and estimating the minimum value of SNR such that the addend 1/SNR can be neglected. This simple guess leads to \footnote{The value of $\epsilon_{1M,{\rm min}}$ at $(\rm SNR)_{\rm high}$ differs from the true asymptotic value only by 1\% for $\ell_{\rm max}=20$ and 5\% for $\ell_{\rm max}=3$.}
\begin{equation}
 (\rm SNR)_{\rm high}\simeq \frac{(\ell_{\rm max}+2)!}{(\ell_{\rm max}-2)!}.
\label{eqn:asympSNR}   
\end{equation}
 The
sensitivity of a PTA to a circular-polarization dipole is
maximized once an SNR of this value is reached.

If we surmise (optimistically) an $\ell_{\rm max} \simeq 20$
(corresponding to $N_p\sim 400$ pulsars), then Eq.~(\ref{eqn:highSNR}) evaluates to
$\epsilon_{1M, {\rm min}}\simeq 2.5$, which is about 1.2
times the largest value, $\epsilon_{1M,{\rm max}} =
\sqrt{4\pi/3}$, that this amplitude can have.  We thus conclude
that an individual component $\epsilon_{1M}$ of the
circular-polarization dipole is not detectable. 

However, if we simply want to establish the existence of a
circular-polarization dipole, without any constraint to its
direction, we will evaluate the overall dipole amplitude,
\begin{equation}
\label{eqn:totaldipole}
     d_c = \left[ \sum_M \left| \epsilon_{1M} \right|^2
     \right]^{1/2}.
\end{equation}
Since this is obtained as the sum, in quadrature, of the three
$\epsilon_{1M}$s, the smallest detectable $d_c$ is about a
factor of $\sqrt{3}$ smaller, implying (with $\ell_{\rm max}
\simeq 20$) that a dipole as small as $0.7$ times the maximal
dipole can actually be detected.
If the local GW background, at the
frequency considered, is dominated by a single source or handful of sources, such a
circular-polarization dipole is certainly conceivable and so worth seeking.
On the other hand, the results discussed so far only hold for very large SNR, such that $\epsilon_{1M,{\rm min}}$ reaches its asymptotic value (for example, when $\ell_{\rm max} \simeq 20$, Eq.~\ref{eqn:asympSNR} gives $(\rm SNR)_{\rm high}\simeq 1.8\times 10^5$). For lower values of SNR there is less room for the observation of the dipole and one can establish numerically a threshold for the overall dipole detection. The corresponding minimal conditions for detection are a number of pulsars $N_p\gtrsim100$ (i.e. $\ell_{\rm max} \gtrsim 10$) and a very clear detection of the isotropic GW background with $\rm SNR\gtrsim 400$. More precisely, for $\ell_{\rm max} = 10$ and $\rm SNR=400$ only an overall dipole equal to $0.98$ times the maximal value can be detected.
\subsection{Other multipoles}

\begin{figure}[h]
\centering
\hspace*{-0.1cm}\includegraphics[width=0.5\textwidth]{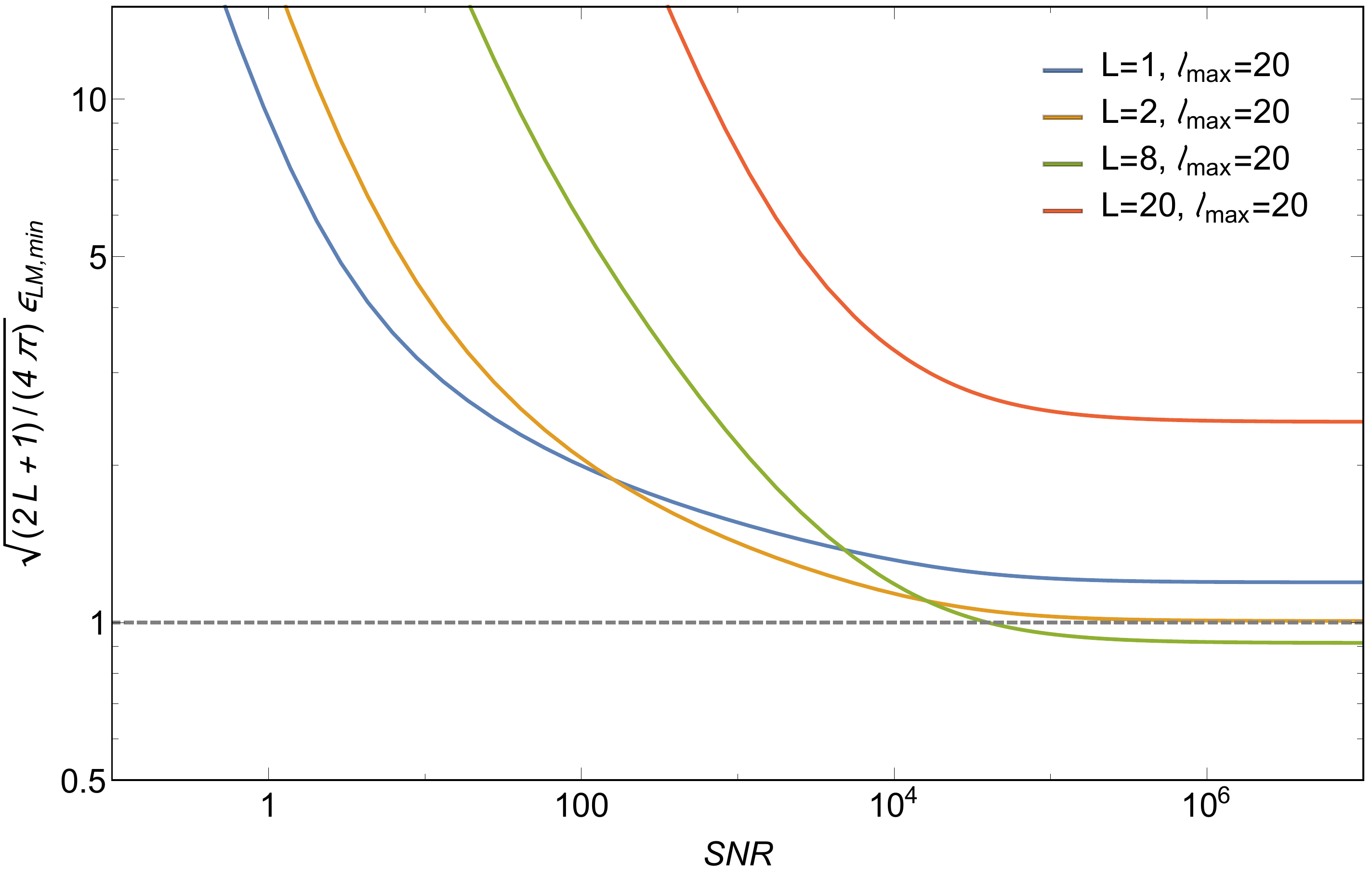}
\caption{The smallest detectable (at the $3\sigma$ level)
circular-polarization multipoles $\epsilon_{LM}$ (normalized to their maximal values $\epsilon_{LM,{\rm max}} =
\sqrt{4\pi/(2L+1)}$), as a function of the SNR with which the isotropic GW background is detected, for several values of $L$ and assuming a maximum timing-residual multipole moment $\ell_{\rm max}=20$.}
\label{fig:multipoles}
\end{figure}

Results for the detectability of higher order multipoles can
be inferred by numerically evaluating the general expression in
Eq.~(\ref{eq:finalvarepsLM}).  
Fig.~\ref{fig:multipoles} shows the smallest detectable multipole coefficients $\epsilon_{LM,{\rm min}}$ (normalized to their maximum possible values $\epsilon_{LM,{\rm max}} =
\sqrt{4\pi/(2L+1)}$) as a function of the isotropic signal SNR, for several values of the multipole $L$ and assuming $\ell_{\rm max}=20$.

The plot and the numerical analyis seem to imply that, for a given $\ell_{\rm max}$, some of the higher-order multipoles have a better detectability than the dipole in the high-SNR regime, because their ratio $\epsilon_{LM,{\rm min}}/\epsilon_{LM,{\rm max}}$ reaches a lower asymptotic value. For $\ell_{\rm max}=20$ the asymptotic value for the minimum quadrupole $\epsilon_{2M,{\rm min}}$ (for any given $M$ component) is just above the maximal ${\epsilon_{2M,{\rm max}}}=\sqrt{4\pi/5}$, while for the overall quadrupole (defined similarly to Eq.~\ref{eqn:totaldipole}) an amount as small as $0.45$ times the maximal quadrupole can be detected. As for the dipole, also higher order multipoles reach the asymptotic regime only at very large SNR, of the order of $(\rm SNR)_{\rm high}$ defined in Eq.~\ref{eqn:asympSNR}. It can be seen from Fig.~\ref{fig:multipoles} that, assuming $\ell_{\rm max}=20$, when $\rm SNR\lesssim100$ (certainly including more realistic values of SNR) the dipole has the lower, and thus the best, value of $\epsilon_{LM,{\rm min}}/\epsilon_{LM,{\rm max}}$ for a single $M$ component, although such SNR is not good enough for single $M$ detections. 
For $\ell_{\rm max}=10$ (corresponding to $N_p=100$) and $\rm SNR=400$, which are the threshold values mentioned at the end of Section~\ref{sec:dipoleresult}, an overall quadrupole equal to $0.72$ times the maximal value is observable, and overall multipoles up to $L=8$ and close to their respective maximal values can also be detected.
Unless the isotropic signal SNR is increased to even more non-realistic values, the conclusions about detectability of dipole and higher multipoles presented here could be altered only with an exponentially large number of pulsars (or, as alluded to below,
by co-adding multiple frequencies in the event that multiple
frequencies have similar SNR).

\section{Conclusions}
\label{sec:conclusions}
We have augmented prior work \cite{Hotinli:2019tpc} to develop
estimators and evaluate the detectability with PTAs of
circular-polarization anisotropies in the stochastic GW
background.  We confirm with this new formalism earlier findings
\cite{Kato:2015bye} that the circular-polarization monopole is
not detectable.  We evaluate the smallest detectable
circular-polarization dipole anisotropy and find that its overall amplitude (i.e. without constraints to the direction) is
conceivably detectable if it is close to maximal, if the
isotropic signal is detected at the $\gtrsim400\sigma$ level, and
at least $N_p\sim100$ pulsars are observed.  In those conditions also a few higher overall multipoles can be detected. 
The results suggest an only logarithmic improvement in the sensitivity with the number
of pulsars. A certain improvement can be obtained with increased overall
signal, but it would be pushed to even more non-realistic values.

We have throughout assumed that the analysis was performed with
just one frequency $f$, whereas in practice there may 
be many frequency channels available.  The analysis can,
however, be done individually for each available frequency and
the results then added in quadrature.  If the stochastic
background is assumed to be uncorrelated at different
frequencies, then the signal-to-noise with which a
circular-polarization anisotropy can be detected will be the
sum, in quadrature, of the signal-to-noise for each individual
channel $f$.  If these signal-to-noises are comparable for all
of the available frequencies, then the SNR could conceivably be
increased by a factor of the square root of the number of
frequencies (a number of order 100 for 10 years of observations
with a two-week cadence).  In practice, though, the signal (the
stochastic GW background) and noise are likely to have different
frequency dependences, and if so, the overall SNR is dominated by
only one, or perhaps a handful, of frequencies.  In this case,
the estimates of the detectability of the circular-polarization
dipole (and higher moments) presented here might be improved,
but probably by no more than a factor of a few.

On the other hand, we have considered an idealization of the measurements in which pulsars are roughly uniformly distributed on the sky and observed with comparable timing-residual noise.  In practice, the distribution is not uniform, and the timing-residual noises vary from one pulsar to another.  These complications are straightforward to deal with using techniques \cite{Kato:2015bye} already developed.  These complications will, however, degrade the sensitivities to circular polarization relative to those obtained with the idealizations adopted here.

Finally, we have focussed here on the PTA characterization of a
stochastic GW background.  There is, however, a close
correspondence between PTA searches and astrometry searches
(see, e.g., Ref.~\cite{Qin:2018yhy}).  Circular-polarization
estimators for astrometry searches should thus be similarly
obtained, and the quantitative conclusions about detectability
similar.  It may also be interesting in future work to
investigate the possibility to co-add information on
circular-polarization and intensity anisotropies that may arise
if the local signal is due to a handful of nearby SMBH pairs.

\begin{acknowledgments}
EB was supported by the
SwissMap National Center for Competence in Research. EB thanks
the Department of Physics and Astronomy at Johns Hopkins
University for hospitality during the development of this work.
MK was supported in part by NASA Grant No.\ NNX17AK38G, NSF Grant
No.\ 1818899, and the Simons Foundation.
\end{acknowledgments}

\end{document}